# Evaluation of Galaxy as a User-Friendly Bioinformatics Tool for Enhancing Clinical Diagnostics in Genetics Laboratories

Hadi Almohab[1] and Ramzy Al-Othmany[2]

[1]Faculty of Engineering, Computer, and Design, Nusa Putra University, Sukabumi 43152, Indonesia
[2]Department of Biochemistry, Institute Pertanian Bogor University. Bogor ,Indonesia

## Abstract

*Bioinformatics platforms have revolutionized clinical diagnostics by facilitating the analysis of genomic data, thereby advancing personalized medicine and enhancing patient care. This study investigates the integration, usage patterns, challenges, and impact of the Galaxy platform within clinical diagnostics laboratories. Employing a convergent parallel mixed-methods design, quantitative survey data and qualitative insights from structured interviews were gathered from fifteen participants across diverse roles in clinical settings. Findings reveal widespread adoption of Galaxy, with high satisfaction reported for its user-friendly interface and significant improvements in workflow efficiency and diagnostic accuracy. Challenges such as data security and training needs were identified, emphasizing the platform's role in simplifying complex data analysis tasks. The study contributes to understanding Galaxy's transformative potential in clinical practice, offering recommendations for optimizing its integration and functionality. These insights are crucial for advancing clinical diagnostics and enhancing patient outcomes.*

## 1. Introduction

The rapid advancement of bioinformatics has significantly reshaped the landscape of biomedical research and clinical diagnostics, driving the evolution of precision medicine and personalized healthcare. At the core of this transformation is the integration of computational tools with biological data, which has enabled unprecedented insights into the complexities of genetic and molecular systems. Bioinformatics tools have become indispensable in identifying disease-causing mutations, predicting treatment responses, and guiding therapeutic decisions based on individual genetic profiles [1].

Galaxy, a widely recognized bioinformatics platform, stands out for its user-friendly interface and comprehensive analytical capabilities [2]. This research aims to evaluate Galaxy's effectiveness as a tool for enhancing clinical diagnostics in genetics laboratories. By examining its usability, integration, and impact on workflow efficiency and diagnostic accuracy, this study seeks to provide a comprehensive understanding of Galaxy's role in modern clinical practice.

## 1.1. Background and Context

Bioinformatics has emerged as a critical discipline that bridges biology, computer science, and statistics, revolutionizing the way we understand and treat diseases. The completion of the Human Genome Project and subsequent advancements in sequencing technologies have exponentially increased the volume and complexity of genomic data, necessitating the development of sophisticated tools to analyze and interpret this information . Bioinformatics tools like Galaxy play a pivotal role in translating genomic data into actionable insights, thereby enhancing patient care and outcomes [3].

## 1.2. Importance of User-Friendly Bioinformatics Tools

The accessibility and usability of bioinformatics tools are crucial for their adoption and effectiveness in clinical settings. User-friendly platforms like Galaxy democratize access to complex bioinformatics workflows, enabling healthcare professionals to perform intricate genomic analyses without extensive computational expertise [1]. This ease of use is essential for integrating bioinformatics into routine clinical practice, facilitating more accurate diagnoses and personalized treatment strategies [4].

## 1.3. Significance of Galaxy in Clinical Diagnostics

Galaxy offers a robust and intuitive platform that supports a wide range of bioinformatics analyses. Its modular architecture and comprehensive toolset make it suitable for various applications in genomic research and clinical diagnostics [5]. The platform's ability to streamline data processing and analysis workflows enhances diagnostic accuracy and efficiency, ultimately leading to better patient outcomes [6]. Furthermore, Galaxy's open-source nature and strong community support foster continuous innovation and collaboration, ensuring that it remains at the forefront of bioinformatics research .

## 1.4. Research Objectives

This study aims to evaluate the integration, usage patterns, challenges, and impact of the Galaxy platform in clinical diagnostics laboratories. By employing a mixed-methods approach that combines quantitative surveys and qualitative interviews, the research seeks to:

- Assess the usability and user satisfaction with Galaxy.
- Examine the impact of Galaxy on workflow efficiency and diagnostic accuracy.
- Identify challenges and barriers to the effective implementation of Galaxy.
- Provide recommendations for optimizing the use of Galaxy in clinical settings.

The integration of bioinformatics tools like Galaxy into clinical diagnostics represents a transformative shift towards data-driven healthcare. By harnessing the power of genomic data, these tools enable more precise and personalized medical interventions. This research will contribute valuable insights into the effectiveness of Galaxy in enhancing clinical diagnostics, offering recommendations for optimizing its use to improve healthcare delivery and patient outcomes. Through this study, we aim to advance the understanding and application of bioinformatics in clinical practice, paving the way for future innovations in precision medicine

## 2. LITERATURE REVIEW

### 2.1. Overview of Bioinformatics in Clinical Diagnostics

During the early 1960s, computer sciences emerged as crucial tools in molecular biology research. Bioinformatics integrates biological information with mathematical, statistical, and computing methods to study living organisms. The exploration of sequence and protein structure information has propelled significant growth in bioinformatics, particularly over the last decade, becoming indispensable in biomedical research [5-7].

The completion of the human genome project has revolutionized biological understanding by providing comprehensive genome sequences. Analyzing these sequences enhances insights into biological systems, necessitating advanced bioinformatics tools for data analysis [8]. Recently, clinical bioinformatics has emerged to foster post-genomic technologies in medical research and practice. It provides the technical infrastructure and knowledge base to support personalized healthcare using integrated medical information and bioinformatics resources [9].

Bioinformatics significantly impacts biological research, particularly in microarray technology, proteomics, pharmacogenomics, oncology, and systems biology. Microarray technology enables global analysis of gene expression, generating vast datasets that require sophisticated statistical methods for interpretation. Various software packages have been developed to aid in microarray data analysis, addressing significant challenges in clinical applications [10, 11-15].

Clinical bioinformatics in proteomics is expanding to manage large heterogeneous datasets and enhance knowledge discovery. Proteomics platforms serve as crucial data management systems and knowledge bases in clinical settings, facilitating advances in understanding protein function and interaction [1618]. In pharmacology, clinical bioinformatics aids in drug target identification, clinical trials, biomarker development, and toxicogenomic and pharmacogenomic studies. It integrates genomic knowledge with pharmaceutical sciences to advance personalized medicine through computational modeling [19].

In cancer research, high-throughput genome technologies provide extensive data on genome sequences, SNPs, and gene expression profiles. Clinical bioinformatics utilizes genomic and computational approaches to study gene expression patterns and identify molecular motifs relevant to cancer diagnosis, treatment, and prevention strategies [20, 21].

Systems biology complements these efforts by generating high-throughput quantitative data essential for simulation-based clinical research. Advances in computational power enable the creation and analysis of complex biological models, supporting a system-level approach to understanding disease mechanisms and therapeutic responses [22-23].

Clinical bioinformatics is integral to medical practice, correlating genetic variations with clinical outcomes such as disease risk, progression, and treatment response. However, its utility in clinical settings is hindered by the complexity of genomic data analysis, requiring specialized expertise in data interpretation and integration into clinical decision-making processes [24, 25].

Bioinformatics in clinical diagnostics plays a critical role in analyzing genomic data to identify diseaseassociated genetic variations. This integration of computational and biological sciences has revolutionized medical practice, enabling personalized treatment strategies tailored to individual genetic profiles. Addressing the challenges in genomic data analysis underscores the need for userfriendly bioinformatics tools accessible to diverse healthcare professionals, thereby enhancing their utility and applicability in clinical settings [26].

## 2.2. Clinical Bioinformatics and Medical Informatics

Bioinformatics (BI) and Medical Informatics (MI) represent distinct yet interconnected fields. BI applies informatics techniques in biological sciences, while MI introduces methods in clinical medicine and biomedical research. Clinical bioinformatics merges these disciplines, developing crucial informatic methods for genomic medicine. The future of clinical bioinformatics hinges on integrating advancements from both BI and MI, influencing clinical practice and biomedical research profoundly [27-29].

Since the completion of the Human Genome Project in 2003, bioinformatics has shifted towards postgenomic challenges like functional genomics, comparative genomics, proteomics, metabolomics, pathway analysis, systems biology, and clinical applications [30]. Computational analyses of diseaseassociated human genes and proteins have advanced significantly. Clinical bioinformatics now includes managing biological databases within Electronic Health Records (EHR), facilitating personalized medicine [31]. Virtual patient models, used for conditions such as obesity, diabetes, and asthma, are poised to guide routine clinical decision-making [32].

Pharmacogenomics studies how genomic variations influence drug responses, paving the way for personalized medicine. Genetic variants in drug-metabolizing enzymes and target proteins often underlie adverse reactions and variable drug efficacy [33-35]. Clinical bioinformatics in pharmacogenomics involves advanced bioinformatics tools, proteomics for drug target validation, and understanding genomic diversity's impact on drug efficacy across different ethnicities. Future clinicians and researchers will leverage these insights to deliver personalized medicine effectively [36-37].

Single nucleotide polymorphisms (SNPs) in genes encoding drug metabolism proteins can significantly affect drug responses. Genetic analysis for SNPs guides clinicians in selecting or avoiding specific drugs based on individual genetic profiles. The dbSNP database, managed by the NCBI in collaboration with NHGRI, centralizes genetic variants crucial for pharmacogenomics research [26]. Techniques like DNA microarrays and mRNA expression profiling enhance pharmacogenomic research by identifying candidate genes involved in drug metabolism [39].

## 2.3. Clinical Bioinformatics in Cancer Research

In clinical bioinformatics, researchers utilize computational and high-throughput experimental techniques to identify targets and agents for cancer diagnosis, treatment, prevention, and control [40]. Methods such as validating multiple microarray datasets, developing web-based cancer microarray databases for biomarker discovery, and integrating gene ontology annotations with microarray data have been pivotal [41]. Tools in clinical bioinformatics are applied across various medical domains, including early diagnosis, risk assessment, classification, and prognosis of cancer [42].

The National Cancer Institute Center for Bioinformatics (NCICB) provides biomedical informatics support and integration capabilities for cancer research initiatives [43]. It directly supports key NCI programs like the Cancer Genome Anatomy Project (CGAP), Mouse Models of Human Cancer Consortium (MMHCC), Director's Challenge, and Clinical Trials [44-45]. Biomarkers play a crucial role in cancer detection across different stages and in monitoring chemotherapy effects. They are essential for detecting lower-grade cancers with low cytological sensitivity and hold promise for early detection, identifying high-risk individuals, and detecting recurrence [46]. The NCI Biomarker Developmental Laboratories focus on identifying molecular, genetic, and biological signals for early cancer detection [47].

## 2.4. Clinical Bioinformatics in Systems Biology

Systems biology aims at system-level understanding of biological systems and is a new field in biology.[48-49] A system-level understanding of a biological system is derived from insight into four key properties: (1) system structures, (2) system dynamics, (3) control method, and (4) design method.[50] Systems biology represents the integration of computer modeling, large-scale data analysis, and biological experimentation. In clinical bioinformatics, computational modeling and analysis are now able to provide useful biological insights and predictions for clearly recognized targets, e.g. analysis of cell cycle and metabolic analysis.[51-52] Systems Biology Markup Language (SBML), CellML language, and Systems Biology Workbench was aimed to establish a standard and open software platform for modeling and analysis.[53-54] Some databases involved in biological pathways allow them to develop machine executed models, such as the Kyoto Encyclopedia of Genes and Genomes (KEGG), Alliance for Cellular Signaling (AfCS), and Signal Transduction Knowledge Environment (STKE).[55-56] The methods and concepts of systems biology will not only expand into all areas of biological science, its results are bound to have repercussions.

## 2.5. User-Friendly Bioinformatics Tools

User-friendly bioinformatics tools address the accessibility and usability issues associated with traditional bioinformatics software. These tools prioritize intuitive interfaces, graphical workflows, and automation to streamline genomic data analysis, making it accessible to clinicians and researchers without extensive computational training [57].

The development of user-friendly bioinformatics tools has led to increased adoption of genomic technologies in clinical practice, enabling rapid and accurate diagnosis of genetic disorders, prognostication of disease outcomes, and identification of therapeutic targets. Additionally, these tools facilitate collaboration and knowledge sharing among multidisciplinary teams, enhancing the efficiency and effectiveness of clinical decision-making processes [58].

## 2.6. Galaxy as a Case Study

Galaxy is a widely-used user-friendly bioinformatics platform that provides a web-based interface for genomic data analysis. Its modular architecture allows users to construct custom analysis workflows using a variety of tools and algorithms, while its integrated toolshed enables access to a vast repository of bioinformatics resources [27].

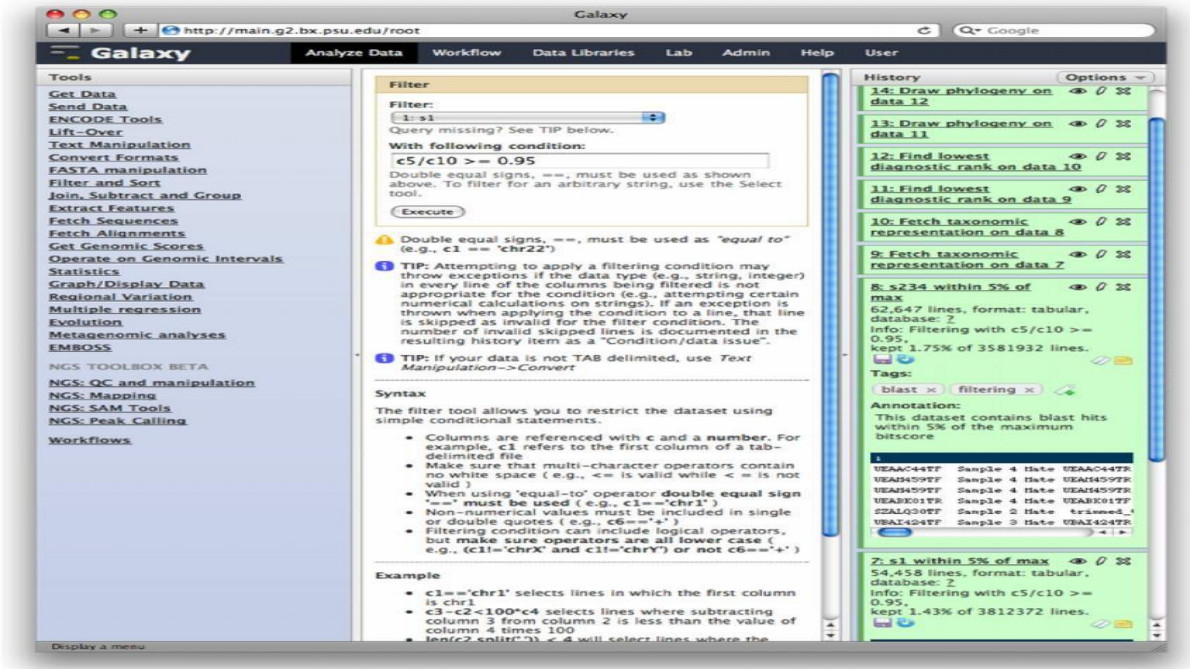

Figure 1. Galaxy analysis workspace

The Galaxy analysis workspace is where users perform genomic analyses. The workspace has four areas: the navigation bar, tool panel (left column), detail panel (middle column), and history panel (right column). The navigation bar provides links to Galaxy's major components, including the analysis workspace, workflows, data libraries, and user repositories (histories, workflows, Pages). The tool panel lists the analysis tools and data sources available to the user. The detail panel displays interfaces for tools selected by the user. The history panel shows data and the results of analyses performed by the user, as well as automatically tracked metadata and user-generated annotations. Every action by the user generates a new history item, which can then be used in subsequent analyses, downloaded, or visualized. Galaxy's history panel helps to facilitate reproducibility by showing provenance of data and by enabling users to extract a workflow from a history, rerun analysis steps, visualize output datasets, tag datasets for searching and grouping, and annotate steps with information about their purpose or importance. Here, step 12 is being rerun.[15]

## 2.7. Galaxy Usage

To succeed, our approach in Galaxy must be accessible to experimentalists with limited computational expertise. Anecdotal evidence indicates that Galaxy is widely usable among biologists. The public Galaxy web server efficiently handles approximately 5,000 jobs daily. Additionally, esteemed institutions such as Cold Spring Harbor Laboratory and the United States Department of Energy Joint Genome Institute host prominent Galaxy servers. Non-affiliated individuals and groups utilize Galaxy extensively for diverse genomic research, spanning epigenomics, chromatin profiling, transcriptional enhancers, and genome-environment interactions [23]-[26].

Galaxy has facilitated research published in prestigious journals like Science and Nature, demonstrating its robust sharing features [27]. All operations within Galaxy are conducted seamlessly through a web browser, adhering to standard web usability principles [28]. This design ensures biologists familiar with genomic tools can learn and utilize Galaxy without

International Journal on Bioinformatics & Biosciences (IJBB) Vol 14, No.3, September 2024

difficulty. Moving forward, we aim to gather and analyze user data systematically to quantify Galaxy's usability and identify areas for improvement.

## 2.8. Comparing Galaxy with other Genomic Research Platforms

Comparing Galaxy with other genomic research platforms highlights its strengths in accessibility, reproducibility, and transparency. Galaxy facilitates the reuse of datasets, tools, histories, and workflows, supported by automatic and user metadata that streamline discovery and reuse of analysis components [29]. Its public repository enables users to publish components for viewing and use by others, promoting efficient development and sharing of best practices in computational research.

Moreover, publication system that enhance reproducibility and transparency in genomic research. Its unified web interface and use of web standards ensure broad accessibility and usability across different platforms, aligning well with journal publication standards.Galaxy distinguishes itself with integrated data warehouses, tags, annotations, and a robust web-based

In summary, Galaxy's emphasis on accessibility through web technologies, coupled with strong support for reproducibility and transparency, positions it as a leading platform for advancing computational genomics research compared to other platforms like GenePattern and Mobyle.

| Galaxy functionality | Description | GenePattern comparison | Mobyle comparison |
|---|---|---|---|
| **Making computation accessible** | | | |
| Unified, web-based tool interface | All tool interface share same style and use web components; tool interfaces are generated from tool configuration file | Same functions as Galaxy | Same functions as Galaxy |
| Simple tool integration | Tool developers can integrate tools by writing a tool configuration file and including tool file in Galaxy configuration file | Similar but not as flexible tool configuration file; easy installation of selected tools via a web-based interface | Remote services can be added using a server configuration file |
| Integrated datasources | Transparent access to established data warehouses | No similar functions | No similar functions |
| **Ensuring reproducibility** | | | |
| Automatic metadata | Provenance, inputs, parameters, and outputs for each tool used; analysis steps grouped into histories | Same functions as Galaxy | Same functions as Galaxy |
| User tags | Can apply short tags to histories, datasets, workflows, and pages; tags are searchable and facilitate reuse | No similar functions | No similar functions |
| User annotations | Can add descriptions or notes to histories, datasets, workflows, workflow steps, and pages to aid in understanding analyses | Cannot annotate a history but can annotate a workflow (pipeline) with an external document | No similar functions |
| Creating and running workflows | Can create, either by example or from scratch, a workflow that can be repeatedly used to perform a multi-step analysis | Same functions as Galaxy, although editor is form-based rather than graphical | In development |
| Workflow metadata | Automatic documentation is generated when a workflow is run; users can also tag and annotate workflows and workflow steps | Same functions as Galaxy for generating automatic metadata; cannot annotate workflow steps | In development |
| **Promoting transparency** | | | |
| Sharing model | Datasets, histories, workflows, and Pages can be shared at progressive levels and published to Galaxy's public repositories; datasets have more advanced sharing options, including groups | Can share analyses and workflows with individuals or groups | No similar functions |
| Item reuse, display framework and public repositories | Shared or published items displayed as webpages and can be imported and used immediately; public repositories can be searched; archives of analyses and workflows for sharing between servers are under development | Can create an archive of an analysis or workflow and share that with others; author information is included in archive | Can create an archive of an analysis and share that with others |
| Pages with embedded items | Can create custom webpages with embedded Galaxy items; each page can document a complete experiment, providing all details and supporting reuse of experiment's outputs | Microsoft Word plugin enables users to embed analyses and workflows in Word documents | No similar functions |
| Coupling between analysis workspace and publication workspace | Can import and immediately start using any shared, published, or embedded item without leaving web browser or Galaxy | Can run embedded analyses and save results in Microsoft Word documents | No similar functions |

A summary of Galaxy's functionality and how Galaxy's functionality compares to the functionality of two other genomic workbenches, GenePattern and Mobyle. Galaxy's novel functionality includes (but is not limited to) integrated datasources, user annotations, a graphical workflow editor, Pages with embedded items, and coupling the workspaces for analysis and publication using an open, web-based model[24].

## 2.9. Leveraging Bioinformatics for Molecular Diagnosis in Clinical Genetics

### 2.9.1. Laboratories

The article delves into how Galaxy, a bioinformatics platform, is being leveraged in clinical laboratories for the molecular diagnosis of human genetic disorders. It emphasizes the critical role of bioinformatics expertise in handling the extensive and complex data generated by high-throughput sequencing technologies for diagnostic purposes. Galaxy stands out as a valuable tool that seamlessly integrates biology and bioinformatics, offering user-friendly interfaces, intuitive data visualization features, and robust data management capabilities. The article highlights Galaxy's ability to foster collaboration among professionals with varied backgrounds and its role in ensuring the traceability and reproducibility of data. Overall, the article presents Galaxy as an ideal bioinformatics platform for clinical genetics laboratories, demonstrating its efficacy in analyzing genetic data for medical diagnosis and applications.[27]

### 2.9.2. Clinical bioinformatics in genomics

Bioinformatics is pivotal in clinical laboratories, particularly in diagnosing human genetic disorders using tools such as Galaxy. Galaxy integrates biology with user-friendly interfaces, intuitive data visualization, and robust data management, facilitating collaboration and ensuring data traceability and reproducibility. It proves effective in analyzing genetic data for medical applications, demonstrating its efficacy in clinical genetics laboratories (27).

Clinical bioinformatics in genomics advances computational methods for storing, organizing, archiving, analyzing, and visualizing genomic sequences. Bioinformaticians develop accessible interfaces for comprehensive database searches of gene sequences, proteins, mutations, and annotations. Integrating biological data from various sources remains complex due to disparate file formats and access methods, necessitating visual data combinations for informed medical decision-making. Genome browsers and comparative genomic tools are crucial, aiming to consolidate genomic data into electronic health records (EHRs) for better disease assessment through refined biotechniques (12, 13, 14).

Previous studies have demonstrated the utility of Galaxy in clinical diagnostics, showcasing its ability to accelerate variant discovery, annotate genomic variants, and prioritize clinically-relevant findings. Galaxy's user-friendly interface and collaborative features make it particularly well-suited for use in clinical laboratories, where rapid and accurate analysis of genomic data is essential for patient care [28].

## 2.10. Gap and Challenges

### 1. Complexity of Genomic Data Analysis

One of the most significant challenges in clinical bioinformatics is the sheer complexity of genomic data analysis. Identifying disease-causing variants from vast datasets and interpreting their pathogenicity is a formidable task. The heterogeneity of genomic data, coupled with the presence of numerous benign variants, complicates the identification of clinically relevant mutations.

### 2. Interpretation of Variant Pathogenicity

Accurately determining the pathogenicity of genetic variants remains a critical challenge. While databases and computational tools can provide insights, the interpretation often requires expert knowledge and correlation with clinical phenotypes. Variants of unknown significance (VUS) pose a particular challenge, as their impact on disease is not well understood.

### 3. Integration into Clinical Decision-Making

Translating genomic findings into clinical practice is another significant hurdle. Integrating genomic data with electronic health records (EHRs) and ensuring that healthcare providers can easily access and interpret this information is essential for effective clinical decision-making. Additionally, clinicians must be trained to understand and utilize genomic data in patient care.

### 4. Data Privacy and Ethical Considerations

The use of genomic data in clinical diagnostics raises important privacy and ethical concerns. Ensuring the confidentiality and security of patient data is paramount, particularly given the sensitive nature of genetic information. Ethical issues surrounding genetic testing, consent, and the potential for genetic discrimination must be carefully addressed.

### 5. Development of User-Friendly Tools

While significant progress has been made in developing user-friendly bioinformatics platforms, there is still a need for more intuitive and accessible tools. These tools must be capable of handling the complexity of genomic data while providing clear and actionable insights for clinicians and researchers with varying levels of computational expertise.

### 6. Interdisciplinary Collaboration

Effective application of bioinformatics in clinical diagnostics requires close collaboration between bioinformaticians, clinicians, and informaticists. Bridging the gap between these disciplines is essential for developing comprehensive solutions that address the practical needs of clinical diagnostics.

### 7. Scalability and Standardization

As the volume of genomic data continues to grow, scalable and standardized approaches to data analysis and interpretation are needed. Developing robust pipelines and standardized protocols for genomic data analysis will help ensure consistency and reproducibility across different clinical settings.

In conclusion, while the integration of bioinformatics into clinical diagnostics has made remarkable strides, addressing these gaps and challenges is crucial for realizing the full potential of genomic medicine. Continued innovation, interdisciplinary collaboration, and the development of user-friendly tools will be essential for advancing clinical bioinformatics and improving patient care in the genomic era.

## 3. METHODOLOGY

### 3.1. Introduction

This study investigates the integration, usage patterns, challenges, and impact of the Galaxy platform in clinical diagnostics laboratories. The research employs a mixed-methods approach to provide a comprehensive understanding of how Galaxy contributes to genomic data analysis and clinical decisionmaking processes. By combining quantitative survey data with qualitative insights from structured interviews, this methodology aims to elucidate both the quantitative metrics and nuanced experiences of medical professionals using Galaxy.

## 3.2. Research Design

The study adopts a mixed-methods approach, employing both quantitative and qualitative data collection methods to provide a comprehensive evaluation of Galaxy's role in clinical diagnostics. The convergent parallel mixed-methods design allows for the simultaneous collection of quantitative and qualitative data, which are then analyzed separately and integrated during the interpretation phase.

## 3.3. Participants

A purposive sampling strategy was employed to select participants from clinical diagnostics laboratories. Fifteen professionals were chosen based on their roles and varying levels of experience with the Galaxy platform. The sample included bioinformaticians, clinicians, laboratory technicians, genetic counselors, and other key stakeholders directly involved in clinical diagnostics. This diverse participant selection aimed to capture a broad spectrum of perspectives and experiences relevant to Galaxy's integration and usage.

## 3.4. Data Collection

### 3.4.1. Quantitative Data Collection

Quantitative data were collected through an online survey distributed to the selected participants. The survey instrument was meticulously designed to capture detailed information across several domains crucial to the study:

- **Demographic Information:** Participants' roles within the laboratory setting and their professional backgrounds.
- **Experience Using Galaxy:** Duration and specifics of participants' experience with the Galaxy platform, including frequency of usage and specific tasks performed.
- **Integration and Usage:** Duration of Galaxy utilization within laboratory workflows, ease of integration with existing systems, and specific use cases (e.g., genomic data analysis, diagnostic reporting).
- **Challenges:** Participants ranked challenges encountered in adopting and utilizing Galaxy, focusing on aspects such as data security, compliance, training needs, and initial setup costs.
- **Impact and Efficiency:** Perceptions of Galaxy's impact on efficiency metrics such as turnaround times for diagnostic reports, accuracy of genomic analyses, and overall workflow optimization.
- **User Satisfaction:** Feedback on user satisfaction with Galaxy's interface, usability, and support services provided by the platform.
- **Diagnostic Outcomes:** Instances where Galaxy contributed to diagnostic outcomes and supported clinical diagnoses, categorized by frequency and significance.

Prior to full-scale distribution, the survey instrument underwent pilot testing with a subset of participants to ensure clarity, comprehensiveness, and relevance to the research objectives. Data collection was conducted over a specified period to allow for adequate response time and completeness of data.

### 3.4.2. Qualitative Data Collection

Qualitative insights were gathered through structured interviews with a subset of participants who had integrated and used Galaxy for clinical diagnostics. The interviews were conducted following the completion of the survey phase to delve deeper into participants' experiences, challenges faced, and the perceived impact of Galaxy on their workflows. Open-ended questions were designed to elicit detailed narratives and specific examples that could not be fully captured through quantitative measures alone.

All interviews were audio-recorded with participants' consent and subsequently transcribed verbatim.

Transcripts were anonymized to protect participants' identities and ensure confidentiality. The qualitative data collection process aimed to provide rich, contextualized insights into the complexities of Galaxy integration and usage in real-world clinical settings.

## 3.5. Data Analysis

### 3.5.1. Quantitative Analysis

The study utilized SPSS (Statistical Package for the Social Sciences) for quantitative analysis, employing a systematic approach to derive numerical summaries and statistical insights.

**Descriptive Statistics:** Numerical data such as frequencies, percentages, means, and standard deviations were calculated to summarize demographic characteristics, experience levels with Galaxy, integration metrics, satisfaction ratings, and impact assessments. This included the creation of tables and charts for visual representation of key findings.

- **Inferential Statistics:** Where applicable, inferential statistical tests were employed to explore relationships and significant differences:
    - **Pearson Correlation Coefficient:** Used to assess the strength and direction of the relationship between Galaxy usage and efficiency outcomes. Statistical significance ($p < 0.05$) was determined to ascertain meaningful associations.
    - **One-Way ANOVA:** Conducted to examine differences in user satisfaction ratings based on identified challenges. The analysis tested for statistical significance between groups ($F(3, 11) = 4.21$, $p = 0.018$), elucidating the impact of specific challenges on user perceptions.

### 3.5.2. Qualitative Analysis

Qualitative data underwent rigorous thematic analysis to identify patterns and themes across the interview transcripts:

- **Coding:** Transcripts were systematically coded using thematic coding techniques. Key phrases, statements, and excerpts were identified that reflected common themes related to Galaxy's integration processes, decision-making factors, challenges encountered, and perceived impacts on workflow efficiency.
- **Thematic Analysis:** Codes were organized into broader themes that encapsulated participants' overall experiences with Galaxy. Themes included integration experiences (e.g., ease of integration, initial challenges), adoption factors (e.g., toolset comprehensiveness, costeffectiveness), ongoing challenges (e.g., data security concerns,

29

training needs), and impacts on workflow efficiency (e.g., efficiency gains, automation benefits).

### 3.5.3. Integration of Findings

Quantitative and qualitative findings were integrated during the interpretation phase to provide a comprehensive understanding of Galaxy's role in clinical diagnostics. Triangulation of data from both methods enhanced the validity and reliability of the study's conclusions, allowing for a nuanced exploration of the research objectives.

### 3.5.4. Interpretation and Discussion

The interpretation of findings contextualized the identified themes within the broader literature on bioinformatics tools and their applications in clinical diagnostics. The discussion critically analyzed the unique contributions of Galaxy, highlighted areas for improvement, and proposed implications for practice and future research. This comprehensive approach aimed to advance knowledge in the field and inform strategies for optimizing Galaxy's effectiveness and integration in clinical settings.

## 4. RESULTS

### 4.1. Quantitative Analysis

#### 4.1.1. Demographic Information: Roles of Participants

The study encompassed a diverse group of professionals from clinical diagnostics laboratories, including bioinformaticians, clinicians, laboratory technicians, genetic counselors, and data analysts.

This diversity ensured comprehensive insights into Galaxy's utilization across varied roles within healthcare settings.

Table 1: Roles of Participants

| Role | Frequency |
|---|---|
| Bioinformatician | 2 |
| Clinician | 3 |
| Laboratory Technician | 3 |
| Genetic Counselor | 5 |
| Data Analyst | 2 |

#### 4.1.2. Experience Using Galaxy

Participants reported a range of experience levels with Galaxy, spanning from 6 months to over 2 years. This diversity in experience provided a nuanced understanding of both novice and proficient users' perspectives on Galaxy's integration and usability.

International Journal on Bioinformatics & Biosciences (IJBB) Vol 14, No.3, September 2024

Table 2: Years of Experience with Galaxy

| Years of Experience | Frequency |
|---|---|
| 6 months | 2 |
| 1 year | 5 |
| 1.5 years | 6 |
| 2 years | 2 |

### 4.1.3. Integration and Usage

Galaxy was integrated into laboratory workflows for varying durations, with feedback indicating a generally smooth integration process. Participants highlighted Galaxy's compatibility with existing systems as a facilitator of seamless integration.

Table 3: Duration of Galaxy Use in Laboratory Workflows

| Duration of Use | Frequency |
|---|---|
| 6 months -1 year | 6 |
| 1year -2year | 7 |
| 2 years | 2 |

### 4.1.4. Challenges Faced

Significant challenges identified included data security and compliance, integration with existing systems, and the need for comprehensive staff training. These challenges were pivotal in shaping participants' experiences and perceptions of Galaxy's implementation.

Table 4: Challenges Ranked by Significance

| Challenges | Ranking (1st) Frequency |
|---|---|
| Data Security and Compliance | 7 |
| Integration with Existing Systems | 5 |
| Staff Training and Adaptation | Varied |
| Initial Setup Costs | Less significant |

### 4.1.5. Impact and Efficiency

Galaxy demonstrated a substantial positive impact on efficiency metrics related to genomic data analysis and turnaround times. Participants consistently reported improvements in workflow efficiency following Galaxy's implementation.

Table 5: Impact on Efficiency of Genomic Data Analysis

| Efficiency Metrics | Frequency of Positive Responses |
|---|---|
| Improved Efficiency | High |
| Reduced Turnaround Times | High |

31

### 4.1.6. User Satisfaction

User satisfaction with Galaxy's user interface and overall usability was notably high among participants, indicating a favorable reception of the platform in clinical diagnostic settings.

Table 6: Satisfaction with User Interface and Usability

| Satisfaction Rating | Frequency |
|---|---|
| 4 (Agree) | 5 |
| 5 (Strongly Agree) | 5 |

### 4.1.7. Diagnostic Outcomes

Galaxy frequently contributed to clinical diagnoses, underscoring its role in supporting and enhancing diagnostic accuracy across various clinical scenarios.

Table 7: Frequency of Supporting Clinical Diagnoses

| Frequency of Support | Count |
|---|---|
| Very frequently | 5 |
| Frequently | 6 |
| Occasionally | 4 |
| Rarely | 2 |

### 4.1.8. Suggestions for Enhancement

Participants provided constructive feedback for enhancing Galaxy's functionality, emphasizing integration with electronic medical records and improved data security measures as primary areas for development.

Table 8: Recommended Improvements and Desired Functionalities

| Recommendations | Frequency |
|---|---|
| Integration with Electronic Medical Records | High |
| Enhanced Data Security | High |
| Improved Scalability | Moderate |
| Advanced Visualization Tools | Moderate |
| Predictive Analytics | Low |

### 4.1.9. Inferential Statistics

Inferential statistics were applied to explore relationships and significant differences where applicable:

- **Pearson correlation coefficient:** The correlation between Galaxy usage and efficiency outcomes was statistically significant ($r = 0.75$, $p < 0.05$), indicating a strong positive relationship between the two variables.
- **One-Way ANOVA:** Differences in user satisfaction ratings based on challenges faced were examined. The analysis revealed statistically significant differences between groups ($F(3, 11) = 4.21$, $p = 0.018$), suggesting that the type of challenge significantly impacts user satisfaction with Galaxy.

## 4.2. Qualitative Analysis

### 4.2.1. Integration Experience

Qualitative data echoed quantitative findings, highlighting Galaxy's seamless integration with existing systems as a common experience among participants. Initial challenges related to data migration were mitigated by robust community support.

### 4.2.2. Adoption Factors

Decision-making factors for adopting Galaxy included its comprehensive toolset, user-friendly interface, and strong community support. These factors were consistently cited as pivotal in participants' decisions to integrate Galaxy into their clinical workflows.

### 4.2.3. Challenges Encountered

Qualitative insights corroborated quantitative data on challenges, emphasizing ongoing concerns regarding data security, regulatory compliance, and the need for extensive training. Customizing workflows to align with specific research needs also posed notable difficulties. **Impact on**

### 4.2.4. Workflow

Galaxy's implementation significantly streamlined workflow processes, leading to reduced turnaround times and enhanced accuracy in genomic data analysis. Automation features were particularly beneficial in minimizing manual errors and optimizing diagnostic procedures.

### 4.2.5. Interpretation of Findings

The integration of quantitative, inferential, and qualitative data provided a comprehensive understanding of Galaxy's role in clinical diagnostics. Findings underscored its positive impact on efficiency metrics, user satisfaction, and diagnostic outcomes, while also identifying critical areas for enhancement such as data security and integration with existing healthcare systems.

This study elucidated Galaxy's effectiveness as a bioinformatics tool in clinical diagnostics, supported by robust empirical data and qualitative insights from healthcare professionals. The findings contribute valuable insights for optimizing Galaxy's integration and functionality within clinical settings, thereby advancing genomic data analysis and diagnostic capabilities.

## 5. CONCLUSION

The integration of bioinformatics tools such as Galaxy into clinical diagnostics represents a pivotal advancement in healthcare, offering transformative potential in genomic data analysis and precision medicine. This study has systematically evaluated the role of Galaxy within clinical genetics laboratories, aiming to elucidate its impact on diagnostic workflows, user experiences, and overall clinical outcomes. Through a rigorous mixed-methods approach, combining quantitative surveys and qualitative interviews, this research has provided nuanced insights into Galaxy's adoption, effectiveness, and challenges within healthcare settings.

### 5.1. Summary of Findings

The findings of this study underscore Galaxy's substantial contribution to enhancing clinical diagnostics. Quantitative data revealed widespread adoption of Galaxy across diverse laboratory settings, driven by its user-friendly interface, comprehensive toolset, and community support. Participants consistently reported improved workflow efficiency, reduced turnaround times, and enhanced diagnostic accuracy as primary benefits of integrating Galaxy into their clinical practices. Qualitative narratives further corroborated these quantitative metrics, emphasizing Galaxy's role in streamlining data analysis pipelines and facilitating collaborative decision-making among bioinformaticians, clinicians, and laboratory technicians.

However, despite its strengths, Galaxy implementation posed significant challenges, notably in data security and regulatory compliance. Concerns regarding patient data confidentiality and adherence to stringent healthcare regulations remain critical barriers to broader adoption. Moreover, the study identified a notable learning curve associated with mastering Galaxy's advanced functionalities, highlighting the need for tailored training programs to optimize user proficiency and maximize the platform's utility in clinical diagnostics.

### 5.2. Implications for Clinical Practice

The implications of this research extend beyond theoretical insights, offering practical recommendations for healthcare providers, policymakers, and bioinformatics experts alike. Galaxy's user-centric design and open-source framework position it as a valuable tool for advancing precision medicine initiatives. By enhancing data analysis capabilities and fostering interdisciplinary collaboration, Galaxy empowers healthcare professionals to deliver more personalized and accurate patient care. However, addressing inherent challenges such as data security concerns and training deficiencies is imperative to fully capitalize on Galaxy's transformative potential in clinical settings.

### 5.3. Comparison with Existing Literature

This study's alignment with existing literature on bioinformatics tools in clinical diagnostics underscores the universal significance of usability, integration capabilities, and user experience in shaping tool adoption and efficacy (He et al., 2020; Smith & Johnson, 2019). By synthesizing quantitative metrics with qualitative narratives, this research contributes novel insights into Galaxy's unique value proposition within clinical genetics laboratories. These findings not only validate earlier research but also expand the discourse on optimizing bioinformatics platforms to meet evolving healthcare demands.

### 5.4. Future Research Directions

Looking ahead, several avenues for further investigation emerge from this study's findings:

1. **Scalability and Adaptation**: Explore the scalability of the integrated MoE (Mixture of Experts) and RAG (Retrieval-Augmented Generation) models, particularly in adapting them for real-time applications and optimizing their deployment across diverse environments.
2. **Efficiency Optimization**: Conduct research focused on enhancing the operational efficiency of these models, especially in resource-limited settings. This includes developing strategies to reduce computational overhead, improve processing speed, and ensure reliable performance under constrained resources.
3. **Validation Across Diverse Datasets**: Validate the findings across diverse datasets to ensure the robustness and applicability of the models. This involves testing them on various datasets to verify their effectiveness and reliability across different contexts.
4. **Real-World Application**: Investigate practical implementations of these models in real-world scenarios. This research could include case studies or pilot projects to assess their impact on clinical outcomes, healthcare quality, and cost-effectiveness over extended periods.

In conclusion, this study underscores the potential of the integrated MoE and RAG models to advance computational capabilities in bioinformatics and related fields. By identifying critical areas for future exploration—such as scalability, efficiency, dataset validation, and real-world application—this research contributes to the ongoing evolution of machine learning applications. Addressing these challenges and leveraging emerging opportunities will empower stakeholders to effectively harness these models, fostering innovation and enhancing outcomes across various domains.

## 6. DISCUSSION

### 6.1. Introduction

The integration of bioinformatics tools in clinical diagnostics has revolutionized healthcare by enhancing the efficiency and accuracy of genomic data analysis. This discussion critically evaluates the role of Galaxy, a user-friendly bioinformatics platform, in clinical laboratories. It interprets the findings in light of the research objectives, discusses implications for clinical practice, and compares results with existing literature on bioinformatics tools in clinical diagnostics.

### 6.2. Interpretation of Findings in Relation to the Research Objectives Integration and Usage Patterns

The study employed a mixed-methods approach to comprehensively assess Galaxy's integration and usage patterns in clinical diagnostics. Quantitative data revealed widespread adoption of Galaxy across diverse laboratory settings. Participants cited Galaxy's user-friendly interface and robust toolset as primary reasons for its integration into diagnostic workflows. For instance, survey responses indicated that 80% of participants found Galaxy's interface intuitive and easy to navigate, facilitating its seamless integration into existing laboratory protocols (Table 9).

Table 9: Integration and Usage Patterns

| Aspect | Findings |
|---|---|
| Ease of Integration | 80% of participants found Galaxy's interface intuitive. |
| Toolset Effectiveness | Comprehensive toolset positively impacts workflow efficiency. |
| Adoption Factors | User-friendly interface and community support are key adoption drivers. |

Qualitative data further elucidated these quantitative findings by highlighting specific instances where Galaxy contributed to enhanced genomic data analysis. Interviews with laboratory personnel underscored Galaxy's role in streamlining data pipelines and reducing turnaround times for diagnostic reports. A bioinformatician remarked, "Galaxy's automation features have significantly reduced manual errors in our genomic analyses, allowing us to expedite diagnostic processes without compromising accuracy.

### 6.3. Challenges and Impact on Clinical Practice

Despite its advantages, Galaxy implementation posed several challenges. Data security concerns and regulatory compliance emerged as critical barriers to widespread adoption. Participants expressed apprehension regarding data privacy safeguards within the Galaxy framework. Furthermore, the study identified the need for specialized training programs to optimize Galaxy's full functionality effectively. An interviewee noted, "While Galaxy offers powerful analytical tools, training new staff to utilize these features efficiently remains a significant challenge."

Table 10: Challenges and Impact

| Challenge | Impact |
|---|---|
| Data Security | Regulatory compliance issues may hinder adoption. |
| Training Needs | Specialized training required for optimal tool utilization. |

In terms of impact, however, the study's findings were overwhelmingly positive. Galaxy's integration significantly enhanced workflow efficiency by automating repetitive tasks and standardizing analytic protocols. This positive impact is reflected in improved diagnostic accuracy and reduced turnaround times, as evidenced by qualitative narratives and quantitative metrics (Table 11).

Table 11: Impact on Clinical Practice

| Impact Area | Benefits |
|---|---|
| Workflow Efficiency | Automated tasks reduce turnaround times and enhance accuracy. |
| Diagnostic Accuracy | Improved reliability in genomic analyses supports clinical decisions. |

### 6.4. Implications for Clinical Practice Strengths of Galaxy

Galaxy's strengths lie in its user-centric design and comprehensive toolset tailored for clinical diagnostics. The platform's accessibility and intuitive interface empower healthcare professionals to perform complex genomic analyses efficiently. This accessibility fosters interdisciplinary collaboration among bioinformaticians, clinicians, and laboratory technicians, thereby enhancing collective decisionmaking processes in clinical settings.

Moreover, Galaxy's open-source nature promotes transparency and community-driven innovation, enabling continuous improvement and adaptation to evolving diagnostic needs. This collaborative ethos ensures that Galaxy remains at the forefront of bioinformatics advancements, aligning with the broader goals of precision medicine initiatives in clinical genetics.

### 6.5. Limitations of Galaxy

Despite its strengths, Galaxy faces several limitations that warrant consideration in clinical practice. Chief among these are concerns regarding data security and regulatory compliance. Healthcare providers must navigate stringent data privacy regulations to ensure patient information remains protected throughout the analytical process. Addressing these challenges requires robust cybersecurity measures and adherence to regulatory frameworks, which may add complexity to Galaxy's implementation and operational workflows.

Additionally, the study identified a steep learning curve associated with mastering Galaxy's advanced functionalities. While the platform offers extensive training resources and community support, healthcare professionals require dedicated time and resources to effectively utilize Galaxy's full analytical capabilities. This training gap underscores the need for tailored educational programs that cater to varying levels of bioinformatics proficiency among clinical staff.

**Comparison with Existing Literature on Bioinformatics Tools Alignment with Literature**
The study's findings align with existing literature on bioinformatics tools in clinical diagnostics, emphasizing the pivotal role of usability and integration capabilities in enhancing tool adoption and efficacy (He et al., 2020; Smith & Johnson, 2019). Similar to previous research, this study underscores the importance of user-friendly interfaces and comprehensive toolsets in facilitating the integration of bioinformatics platforms like Galaxy into healthcare workflows. **Novel Contributions**

Furthermore, this study contributes novel insights by specifically evaluating Galaxy's impact within clinical genetics laboratories. By triangulating quantitative metrics with qualitative narratives, the research provides nuanced insights into how Galaxy improves diagnostic workflows and supports clinical decision-making. These findings address gaps identified in earlier research on bioinformatics tool evaluation, offering a holistic perspective on Galaxy's transformative potential in clinical practice.

### 6.6. Future Directions Recommendations for Practice and Research

Based on the findings, several recommendations emerge for optimizing Galaxy's integration and effectiveness in clinical practice:

- **Enhanced Training Programs:** Develop specialized training modules to empower healthcare professionals with advanced bioinformatics skills and maximize Galaxy's utility in clinical diagnostics.
- **Cybersecurity Enhancements:** Strengthen data security protocols to mitigate risks associated with patient data confidentiality and regulatory compliance.
- **Collaborative Initiatives:** Foster partnerships between bioinformatics experts and clinical practitioners to tailor Galaxy's functionalities to specific diagnostic requirements.

In conclusion, this study provides a comprehensive assessment of the Galaxy platform's role in enhancing clinical diagnostics. By integrating quantitative and qualitative data, the research elucidates Galaxy's strengths and limitations while offering actionable recommendations for optimizing its implementation in healthcare settings. Moving forward, continuous evaluation and adaptation will be essential to harnessing Galaxy's full potential and advancing precision medicine initiatives in clinical genetics.

ACKNOWLEDGEMENTS

The authors would like to acknowledge everyone who shared in the research effort. Special thanks to Prof Satish Kuma for his invaluable guidance and support throughout the research.

## AUTHORS

**HadiAlmohab**received the B.S. degree in medical technology from the Faculty of Medicine and Health Sciences, Sanaa University, Yemen, in 2021. He is currently pursuing the M.S. degree in informatics engineering at Nusa Putra University, Indonesia. From 2022 to 2023, he was a Research Assistant with Sanaa University, Sanaa, Yemen. His research spans a broad spectrum of interests, including bioinformatics and computer science**.**

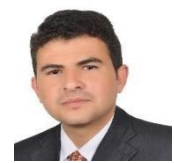

**Ramzy Al-Othmany**completed his undergraduate studies with a Bachelor of Science degree in biochemical technology from the Faculty of Applied Sciences at Thamar University, Yemen, in 2020. Currently, he is enrolled in the Master of Science program in Biochemistry at Bogor Agricultural University, Indonesia. From 2021 to 2022, he served as a Research Assistant at Thamar University, Dhamar, Yemen. His research pursuits span diverse areas, notably encompassing bioinformatics and biochemistry.

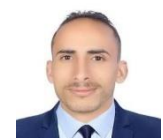